\documentclass[12pt,preprint]{aastex}
\newcommand{\msun}{M\ensuremath{_\odot}}	
\newcommand{\mdot}{\ensuremath{\dot M}}		
\newcommand{\about}{\ensuremath{\sim}}		
\newcommand{\cmd}{color-magnitude diagram}	
\newcommand{\bb}{blackbody}			
\newcommand{\ns}{neutron star}			
\newcommand{\bh}{black hole}			
\newcommand{\xrt}{X-ray transient}      	
\newcommand{\sxt}{soft X-ray transient}      	
\newcommand{\aql}{Aquila~X-1}

\shorttitle{OIR Observations of \aql}
\shortauthors{Maitra \& Bailyn}

\begin{document}
\title{Outburst Morphology in the Soft X-ray Transient \aql}

\author{Dipankar Maitra}
\affil{Sterrenkundig Instituut ``Anton Pannekoek'', University of Amsterdam,
	Kruislaan 403, 1098 SJ Amsterdam, The Netherlands}
\email{D.Maitra@uva.nl}
\and

\author{Charles D. Bailyn}
\affil{Yale University, Department of Astronomy, P.O.Box 208101, 
	New Haven, CT 06520-8101, USA}
\email{charles.bailyn@yale.edu}

\begin{abstract}
We present optical and near-IR (OIR) observations of the major outbursts 
of the \ns\ soft \xrt\ binary system \aql, from summer 1998 -- fall 2007. 
The major outbursts of the source over the observed timespan seem to 
exhibit two main types of light curve morphologies, (a) the classical {\em 
Fast-Rise and Exponential-Decay} (FRED) type outburst seen in many soft 
X-ray transients and (b) the {\em Low-Intensity State} (LIS) where the 
optical-to-soft-X-ray flux ratio is much higher than that seen during a 
FRED. Thus there is no single correlation between the optical (R-band) and 
soft X-ray (1.5-12 keV, as seen by the ASM onboard RXTE) fluxes even 
within the hard state for \aql, suggesting that LISs and FREDs have 
fundamentally different accretion flow properties. Time evolution of the 
OIR fluxes during the major LIS and FRED outbursts is compatible with 
thermal heating of the irradiated outer accretion disk. No signature of 
X-ray spectral state changes or any compact jet are seen in the OIR, 
showing that the OIR \cmd\ (CMD) can be used as a diagnostic tool to 
separate thermal and non-thermal radiation from X-ray binaries where 
orbital and physical parameters of the system are reasonably well known. 
We suggest that the LIS may be caused by truncation of the inner disk in a 
relatively high \mdot\ state, possibly due to matter being diverted into a 
weak outflow. 
\end{abstract}

\keywords{accretion, accretion disks --- stars: neutron --- stars: --- 
X-rays: binaries --- individual (Aquila~X-1)}

\section{Introduction\label{sec:introduction}}

\aql\ is one of the most frequently recurring soft X-ray transients in the 
sky. It goes into outburst approximately once every year although the 
periodicity is not fixed \citep{simon2002}. The frequent outburst 
recurrence makes it an ideal source for studying accretion flow during an 
outburst as well as changes in accretion flow from outburst to outburst. 
The physical and orbital parameters of this system are relatively well 
known. The orbital period of the binary system was determined to be 
$18.71\pm0.06$ hours by \citet{wry2000} from optical photometry during 
quiescence. \citet{c1999} estimated the quiescent V-band magnitude to be 
$21.6\pm0.10$ magnitude. However, there is a contaminating star about 
0.48\arcsec\ East of \aql\ which has a V magnitude of $19.42\pm0.06$ 
\citep{c1999}. The distance to the system is between 4-6.5 kpc as inferred 
both from (a) limits on optical brightness of the Roche lobe filling donor 
star of spectral type between K6-M0 \citep{c1999,r2001} and (b) peak X-ray 
fluxes during type-1 bursts \citep{ccg1987,r2001}.

In systems like \aql\ and other Roche lobe overflowing \sxt\ (SXT) binary 
systems, the thermal viscous disk instability model \citep[DIM;][]{l2001} 
is usually invoked to explain the typical {\em fast-rise and 
exponential-decay} (FRED) type outburst light curve morphology 
\citep{csl1997}. The DIM alone however cannot produce realistic FRED 
outbursts. Another mechanism that plays a key role during outbursts of 
SXTs is X-ray irradiation. The irradiation is caused by the intense X-rays 
from the inner regions of the accretion disk and \ns\ surface (if the 
system harbors a \ns) impinging on the outer disk/donor, and by 
re-emission of this impinged energy in lower energies 
\citep{c1976,v1990,vPM1994,m1995,d1999,obrien2002,h2002,r2006}. Another 
catagory of modifications to the DIM is generated by the possible 
truncation of the inner region of the accretion flow, where the disk 
inflow can be replaced by a radiatively inefficient accretion flow (RIAF).  
There is an increasing body of work on this topic 
\citep{i1977,r1982,ny1994,bb1999,m2001}, but the overall impact of these 
efforts on the outburst morphology is currently still unclear.

Collimated relativistic outflows, loosely termed as ``jets'', have been 
observed in several \sxt s \citep[see][for a review]{f2006} and their 
broad band spectral energy distribution (SED) from radio to X-rays have 
been fitted with jet+disk models comprising of a freely 
expanding ``Blandford - K\"onigl'' jet \citep{bk1979,fb1999,mnw2005} and a 
thermal-viscous ``Shakura-Sunyaev'' accretion disk \citep{ss1973}. 
Observations suggest that compared to black hole (BH) systems, the neutron 
star (NS) SXTs tend to have intrinsically weaker jets and also the 
transition from optically-thick to optically-thin jet spectrum in NS 
systems probably occurs in the mid-IR \citep[][this 
work]{mf2006,migliari2006}, compared to that in near-IR for BH systems 
\citep{cf2002,bb2004,h2005,r2006}.

During outburst, light curve morphology of many sources are not FREDs. 
The X-ray spectral and temporal features also show an enormous diversity
during outbursts of SXT systems, and are categorized into various {\em
states}. 
See e.g. \citet{mr2006} and \citet{hb2005} for extensive discussions on 
definition of states and \citet{vdk2006} for their timing properties.
Despite the multitude of states observed in the different systems and 
different outbursts, two ``canonical'' states can be identified during the 
outburst of most SXTs, characterised by presence or absence of a thermal and 
nonthermal component in the energy spectrum. In the temporal domain, presence 
or absence of a thermal component usually shows up respectively as low or 
high rms variability in the light curve. A host of other intermediate and 
extreme states are also seen, which could be a superposition of the thermal 
and nonthermal components with varying contributions. Precise 
determination of X-ray state requires modeling of spectral energy distribution
and Fourier power density spectra. Since this work primarily focusses mainly 
on OIR observations, we have used dates of X-ray state transitions of \aql\ 
from published studies of X-ray spectral and temporal properties 
of the source \citep{mc2003a,mc2003b,rks2004,mb2004,rodriguez2006,y2007} to
study correlation between X-ray state and OIR flux-density.

We 
have observed \aql\ for $\sim9$ years, through many outbursts, to explore 
the nature of outburst morphologies in a single object. Even though the 
OIR (as well as X-ray) light curve morphologies from outburst to outburst 
differ, we find systematic and repeated tracks traced by the source in OIR 
\cmd s. Since we have long-term data in two bandpasses (R and J) only, we 
have used a simple model to show that the color evolution and the 
resulting track are consistent with heating of a constant area blackbody 
and can be caused by an irradiated outer disc.

In \S\ref{sec:data} we describe the general reduction procedures adopted 
in this work. The light curve morphologies in J, R and ASM bands (of the 
RXTE satellite) and optical--soft X-ray correlations during the major 
outbursts are described in \S\ref{sec:ltc}. Evolution of \aql\ on the OIR 
\cmd\ is presented in \S\ref{sec:colors}. Finally we present our 
discussions in \S\ref{sec:discuss}.

\section{Observations and Data Analysis\label{sec:data}}
\subsection{Optical and Near-IR data\label{sec:oir-data}} 

Our long term monitoring campaign used the Yale 1.0m telescope at Cerro 
Tololo Inter-American Observatory (CTIO) in Chile to observe \aql\ from 
1998-2002 on a daily basis (weather and Sun-angle constraints permitting).  
Since February 2003, the CTIO 1.3m telescope has been used for the 
monitoring. In both cases, images were obtained using the 
ANDICAM\footnote{http://www.astro.yale.edu/smarts/ANDICAM} instrument, 
which is a dual-channel imager capable of simultaneously recording images 
at an optical and an infrared wavelength. Prior to MJD 52430 (2002 June 
5), the optical R-band data were recorded by a Lick/Loral-3 
$2048\times2048$ CCD.  Thereafter, the R-band has been recorded by a 
Fairchild 447 $2048\times2048$ CCD. The near-IR J-band data are recorded 
by a Rockwell $1024\times1024$ HgCdTe Astronomical Wide Area Infrared 
Imager (HAWAII) imager. Between MJD 51130--51367 (1998 November 13 -- 1999 
July 8th), the source was observed using a ``wide-R'' filter. The wide-R 
filter has a broader bandpass than the Johnson R filter used during the 
rest of the campaign. Even though a determination of the Johnson R-band 
flux cannot be made from the wide-R-band measurements, the light curve 
morphology changes little due to the change of filter. We calculated the 
quiescent wide-R-band differential magnitude (between MJD 51232 and MJD 
51306) and applied an offset to the differential magnitudes obtained using 
wide-R filter so that the quiescent level was set to 18.8 magnitude, which 
is the quiescent R-band magnitude of \aql.

Under normal viewing conditions at CTIO, it is extremely difficult to 
separate the true optical counterpart of \aql\ from the bright contaminating star. 
Therefore the magnitudes reported in this work are the combined magnitudes 
of the \aql\ system and the contaminating star, and when we say magnitude of \aql\ 
we actually mean the combined magnitude of \aql\ and the contaminating star. In 
quiescence, light from the contaminating star dominates, but once an outburst 
begins, photons from \aql\ dominate.

The optical data reduction, including bias subtraction, overscan 
correction, and flat-fielding, was done through the standard IRAF data 
reduction pipeline developed for these instruments\footnote{For details 
see http://www.astro.yale.edu/smarts/smarts13m/optprocessing.txt}. For the 
J-band near-IR data, multiple dithered frames were taken and then 
flat-fielded, sky subtracted, aligned and average-combined using an 
in-house IRAF script. During a large portion of the year 2004, one of the 
four quadrants of the IR CCD was not working. In most of the observations 
during this period, one or more of the reference stars were not present in 
the frames. We used whichever comparison stars were present and adjusted 
the resulting differential magnitudes appropriately.

We used the field stars 2MASS 19111739+0035039 and 2MASS 19111633+0034333 
with J-band magnitudes of 12.404 and 12.490 respectively\footnote{Taken 
from the 2MASS point source catalogue at 
http://www.ipac.caltech.edu/2mass}, to obtain differential magnitude of 
the source. Flux calibration of these reference stars in the R-band was 
performed by comparing with Landolt standard stars on the photometric 
night of MJD 53904 (2006 June 18). In order to estimate time variability 
in flux from any of the reference stars, we calculated the difference 
between the observed magnitudes of the reference stars on every night. 
Smaller than 0.02 mag in R-band, this difference is the primary source of 
error in flux determination because during outbursts the Poisson error due 
to photon count-rate is much smaller. The $S/N$ is even better in J-band 
where interstellar extinction is lesser.

For magnitude to flux conversion, following \citet{bessell1998} and
\citet{c1985}, we 
assumed that R=0 magnitude converts to 3064 Jy and J=0 magnitude converts to 
1603 Jy. \citet{wry2000} estimated that $83\pm2 \%$ of the quiescent 
R-band flux is from the contaminating star. Assuming that the brightness of the 
contaminating star is constant in time, we subtracted this flux from the total 
observed flux to obtain the flux of \aql\ only, during quiescence as well 
as outbursts. Interstellar reddening and extinction was calculated 
assuming a hydrogen column density ($N_H$) of $3.4\times 10^{21}$ 
atoms/cm$^2$ in the direction of \aql\ \citep{dl1990}. From deep
optical photometry and spectroscopy of the source during quiescence 
\citet{c1999} estimated the color excess $E(B-V)=0.5\pm0.1$ which is 
consistent with the \citet{dl1990} value of $N_H$ in the direction of
\aql. For dereddening the OIR flux we used the standard $N_H$ to 
visual extinction ($A_V$) correlation given by $N_H/A_V=1.79\times10^{21}$ 
atoms/cm$^2$/mag \citep{ps1995} and the extinction law of \citet{ccm1989} 
which gives $A_R/A_V=0.748$ magnitude and $A_J/A_V=0.282$ magnitude 
for \aql. Taking reddening and extinction in the direction of \aql\ into 
account, the observed $R-J$ color is linearly related to the corresponding
intrinsic spectral index ($\alpha_{RJ;int}$) by the relation
\begin{equation}
 \alpha_{RJ;int} = \frac{log (f_{R}) - log (f_{J})}
                        {log (\nu_R) - log (\nu_J)}
 = 2.47 - 1.63\times (R-J)_{observed}.
 \label{eq:color-slope}
\end{equation}

\subsection{X-ray data\label{sec:asm-data}} 

The X-ray data was taken from the All Sky Monitor \citep[ASM;][]{l1996} 
aboard the Rossi X-ray Timing Explorer \citep[RXTE;][]{j1996} satellite. 
The ASM has been monitoring the sky in the 1.5-3, 3-5 and 5-12 keV energy 
bands continuously since 1996. Usually a given source is observed a few 
times every day, subject to constraints like angular proximity to the Sun 
etc.  The results are made publicly available by the MIT/ASM team through 
their website\footnote{http://xte.mit.edu}. We have used the one day 
average light curve for \aql.  During quiescence, the flux from \aql\ is 
below the detection limit of the ASM, however since we are interested only 
in the major outbursts, which are detectable by the ASM, we did not take 
into account data during quiescence, for our optical-X-ray correlation 
study. Also, during the early as well as the late stages of the outbursts, 
the count-rates in the individual energy channels are small; therefore we 
have only looked at the summed count-rates in this work. 
The ASM count-rates were converted to flux-density at 5.2 keV, assuming a 
Crab-like spectrum.  The Crab spectrum in the ASM energy range is a simple 
power law with a photon-index of -2.1 and the total flux within that range 
is $2.8\times10^{-8}$ ergs/cm$^2$/s \citep{s2002}, so the flux-density at 
5.2 keV is 1.06 mJy. The flux conversion (from ASM count-rate to $\mu$Jy) is 
therefore 
a numerical factor by which the total ASM count rate is multiplied.
The choice of 5.2 keV was motivated by the fact that it is the geometric mean
of the half-transmission points of the ASM response shape\footnote{Taken from 
appendix F of the RXTE technical appendix 
(http://heasarc.nasa.gov/docs/xte/appendix\_f.html) and Ron Remillard, 
priv. comm.} near 2.3 and 12 keV. 
We used geometric mean (GM) because the GM compensates for the fact that 
almost all X-ray spectra slope downward in count rate toward higher energies.

Evolution of spectral shape during an outburst introduces uncertainities in
flux determination. 
Pivoting uncertainities (due to spectral evolution) in ASM's X-ray range 
are small compared to the optical to X-ray separation in frequency 
space. 
We used the online 
{\em WebPIMMS}\footnote{http://heasarc.gsfc.nasa.gov/Tools/w3pimms.html} tool
which shows that for a given total ASM count-rate, the difference in flux
in the ASM band-pass is less than 10\% between a power law of photon index 
2.1 (i.e. Crab spectrum), a power law of photon index 1.7 or a blackbody at 
2 keV \citep[typical photon index/blackbody temperature of \aql\ in hard/soft 
state; see e.g.][]{mc2003b,mb2004,rodriguez2006}.
Therefore we added
a 10\% systematic error, in quadrature to the Poisson error, during the ASM 
count-rate to flux conversion to account for the uncertainity in the spectral
shape.
The precise value for pinning and rescaling the flux conversion
is not very sensitive to the choice of keV reference, as long as this is 
done self-consistently (i.e. scale to Crab flux density at the reference keV, 
and implemented in this work).

In this process, any information about 
changes in the energy spectrum is lost, but we are primarily concerned 
with overall X-ray flux here, and using all the counts results in better 
signal-to-noise ratio. In order to assess the X-ray state of the source, we
have used the dates of state transitions from published literature
\citet{mc2003a,mc2003b,rks2004,mb2004,rodriguez2006,y2007}. For the most recent 
outbursts (e.g. outburst \#9, \#10) with no published dates of state 
transition(s), 
we used the (9.7-16 keV/6-9.7 keV) hardness ratio from archival RXTE/PCA 
pointed observations (Manuel Linares, priv. comm.) whose value correlates 
strongly with the X-ray state of the source. For outbursts with sparse 
coverage of pointed observations (e.g. outburst \#5) we used the 
(5-12 keV/3-5 keV) hardness ratio from the RXTE/ASM data to infer the spectral
state.

\section{Major Outbursts of \aql\ since 1998 June 20\label{sec:ltc}} 

\begin{figure}
\begin{center}
\plotone{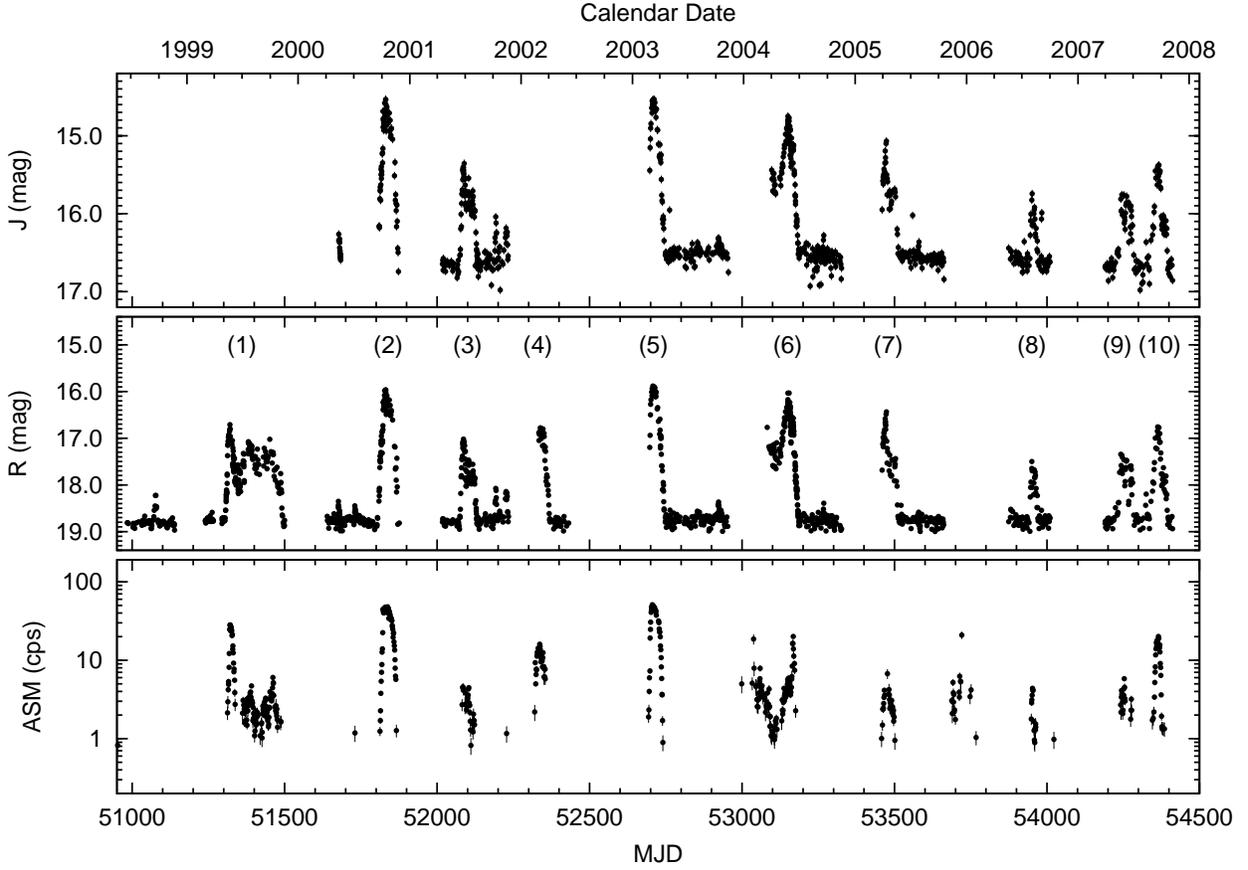}
\caption[Multi-wavelength long-term monitoring of \aql]
{Multi-wavelength long-term monitoring of \aql\ between 
MJD 50984 -- MJD 54411 (1998 June 20 -- 2007 Nov 7).
Top panel: Near-IR J-band light curve. 
Middle panel: Optical R-band light curve. The numbers in
parentheses refer to the serial number assigned to the major outbursts in 
Table ~\ref{tab:outburst_log}.
Bottom panel: Logarithm of soft X-ray (\about 1.5-12 keV) count-rate as 
measured by the All-Sky Monitor onboard the RXTE. Since the ASM does not
detect the source when in quiescence, only those ASM data are plotted for which
the signal-to-noise ratio is $>4$.
The R and J-band fluxes include the flux from \aql\ as well as from
the contaminating star. During periods of quiescence, the soft X-ray flux is below the
detection limit of ASM but the combined flux from \aql+contaminating star remains 
detectable in both R and J bands. Note that a major outburst which occurred 
during late 2005 -- early 2006 could be seen only with the ASM but not the
ground based telescopes due to close angular proximity to the Sun. Therefore
we have not labelled this outburst in the middle panel.
\label{longterm}}
\end{center}
\end{figure}

In Fig.~\ref{longterm} we show the results of long-term multi-wavelength 
monitoring of \aql. The RXTE/ASM monitoring of \aql\ dates back to the 
beginning of 1996, but our nightly R-band monitoring started on MJD 50984 
(1998 June 20th). The J-band observations started MJD 51677 (2000 May 
13th) onwards, but the coverage was somewhat sporadic till the end of 
2002, as can be seen from the top panel of Fig.~\ref{longterm}. 
During quiescent periods, the system is detected in both of the OIR bands but 
not in ASM. Therefore in Fig.~\ref{longterm} we have only plotted those ASM
points for which the signal-to-noise ratio is better than four.

Since observing constraints sometimes make it difficult to date the 
precise starting and ending moments of an outburst, in 
Table~\ref{tab:outburst_log} we note the MJD when the ASM, R-band and 
J-band fluxes attained a maximum during an outburst. We define a major 
outburst as a period of activity of the source, spanning at least 20 days 
or longer, when the source flux was at least $3\sigma$ above quiescence in 
at least one of the observed (J, R and ASM) bandpasses. Eleven major 
periods of activity were observed by the ASM during this period. All these 
periods of major activity were also seen in OIR, except the one during 
late 2005--early 2006 when the source's angular proximity to the Sun was 
too close to observe with ground based OIR telescopes. Even the RXTE/ASM 
data, while showing unmistakeable signs of activity during this particular 
outburst, does not have sufficient coverage and $S/N$ to classify the light 
curve morphology.

Besides the major outbursts, several smaller and shorter periods of 
activity are seen in the OIR light curve (Fig.~\ref{longterm}). Because of 
the low signal-to-noise ratio, particularly in ASM data, and scarcity of 
PCA pointings during these periods of minor activity, they have not been 
discussed in the present work. However, the time of peak OIR flux from the 
source during these ``mini'' outbursts have been presented in 
Table~\ref{tab:outburst_log}.

\subsection{Lightcurve morphology: FREDs and LISs}

\begin{figure}
\begin{center}
\includegraphics{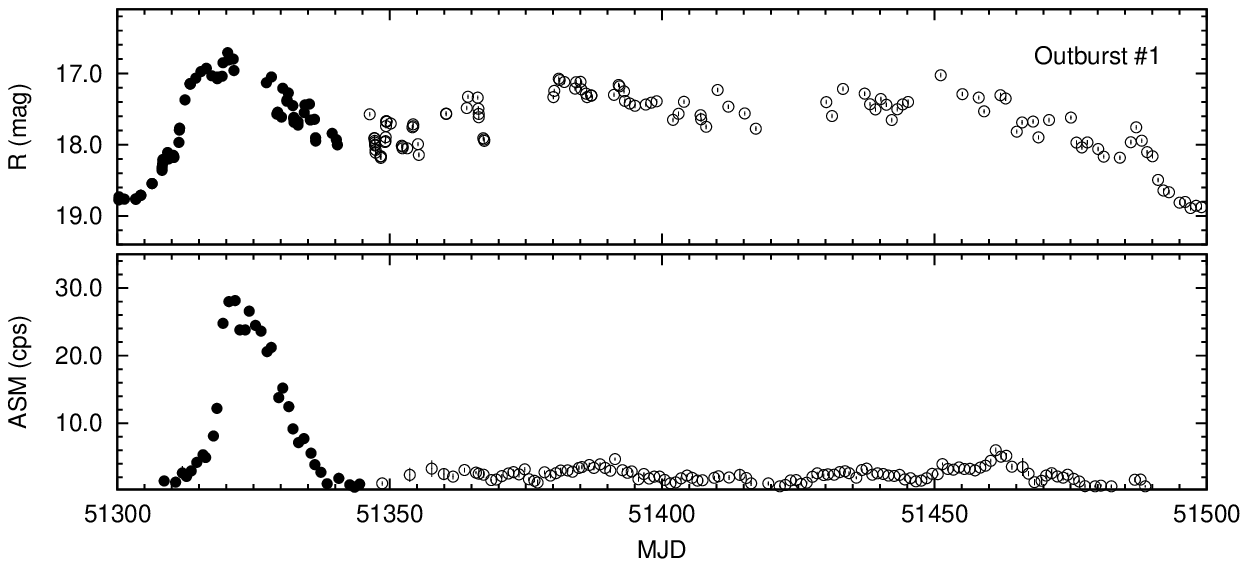}
\includegraphics{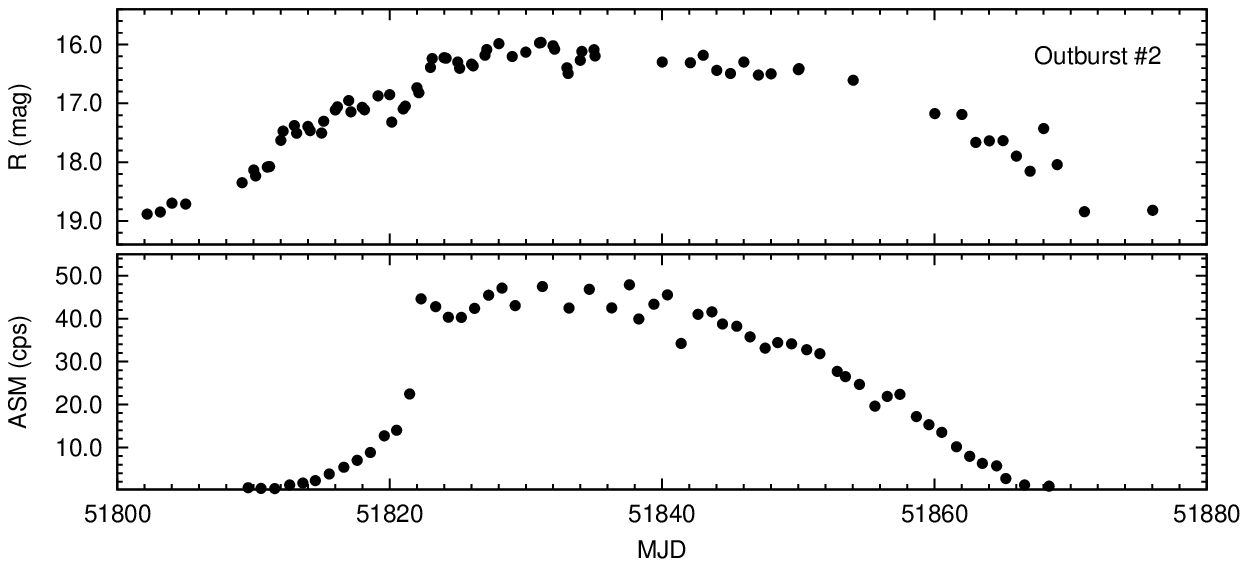}
\includegraphics{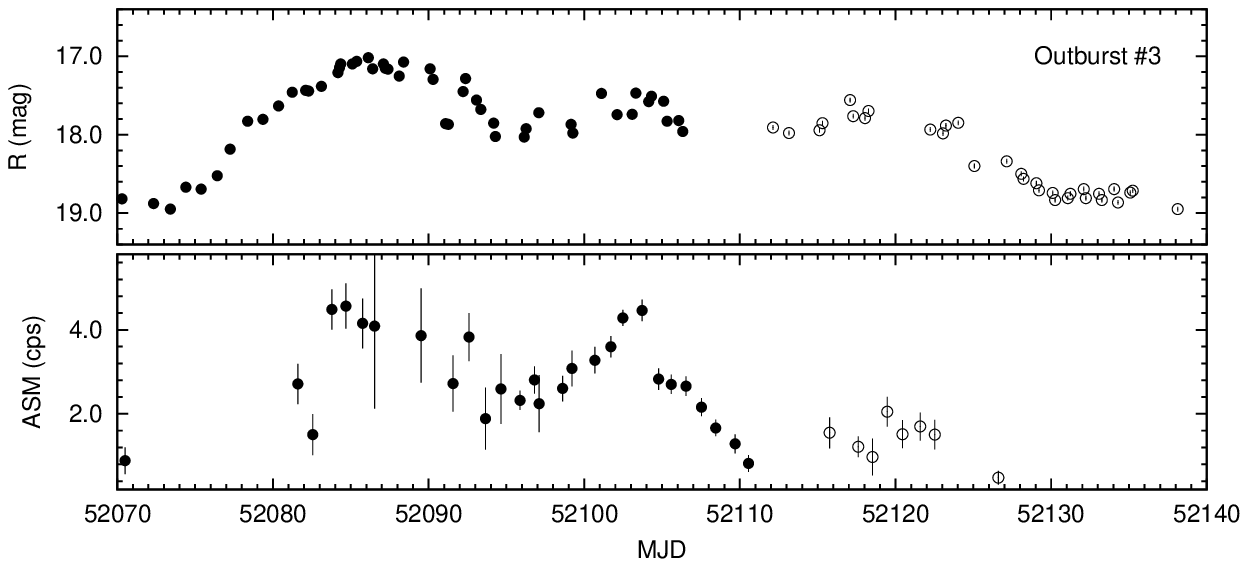}
\caption[Light curve during outbursts of \aql]
{Zooming into the R-band and ASM light curves shown in Fig.~\ref{longterm},
showing the detailed light curve morphology during major outbursts 1--3. 
FRED outbursts are marked by filled circles, LIS by open circles. 
In this figure we are using a linear scale for the ASM count-rates to 
emphasize the difference between LIS and FRED.
\label{zoom_lc_1-3}}
\end{center}
\end{figure}

\begin{figure}
\begin{center}
\includegraphics{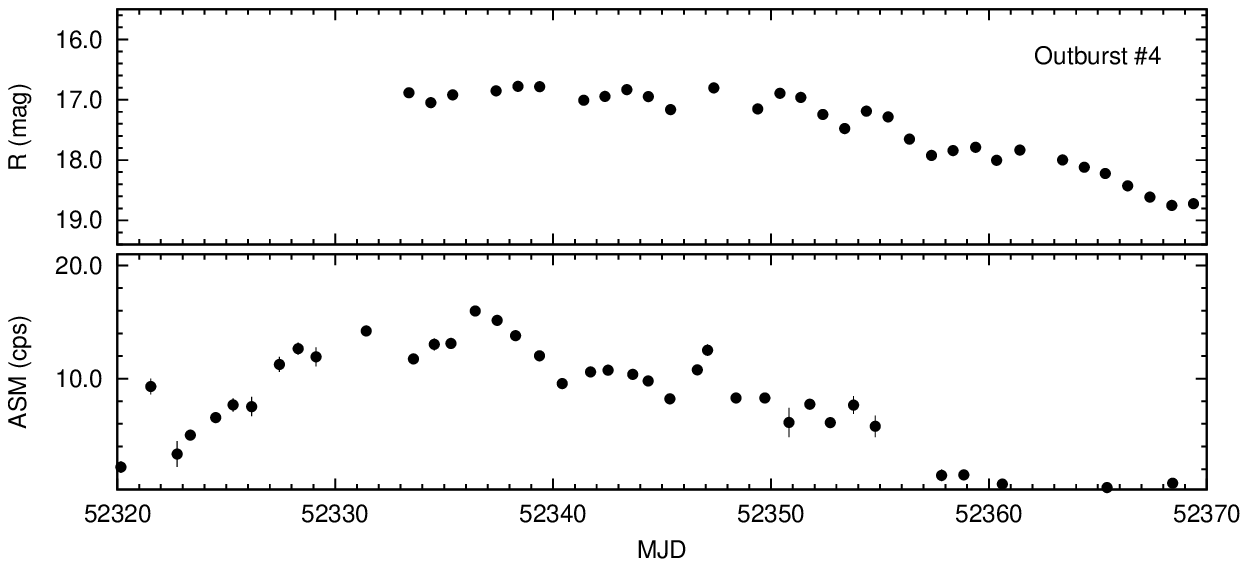}
\includegraphics{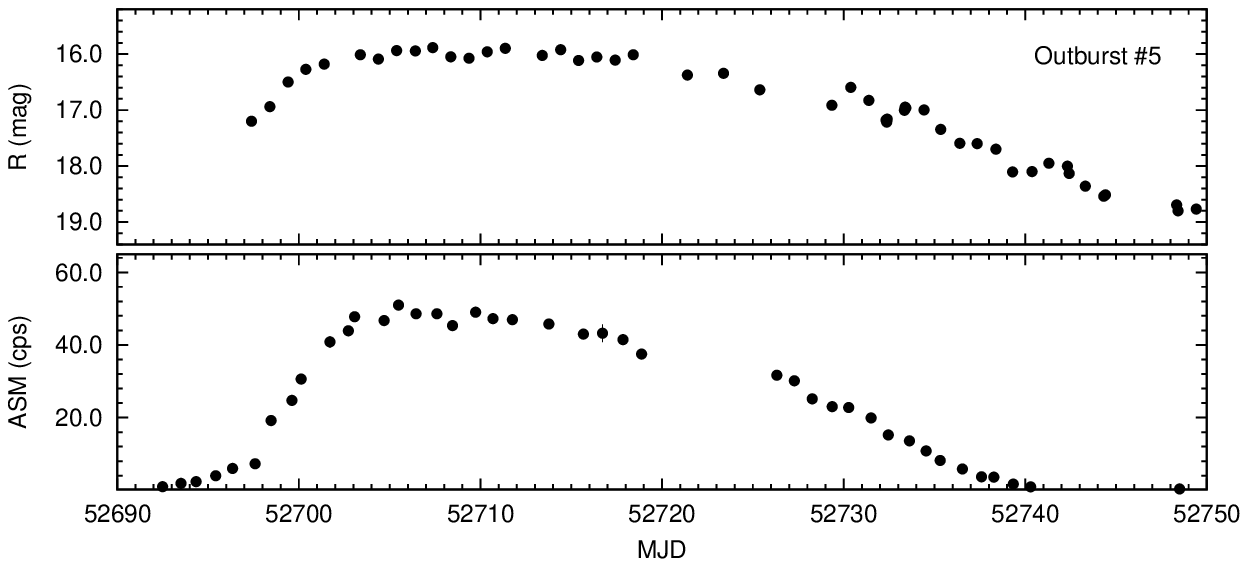}
\includegraphics{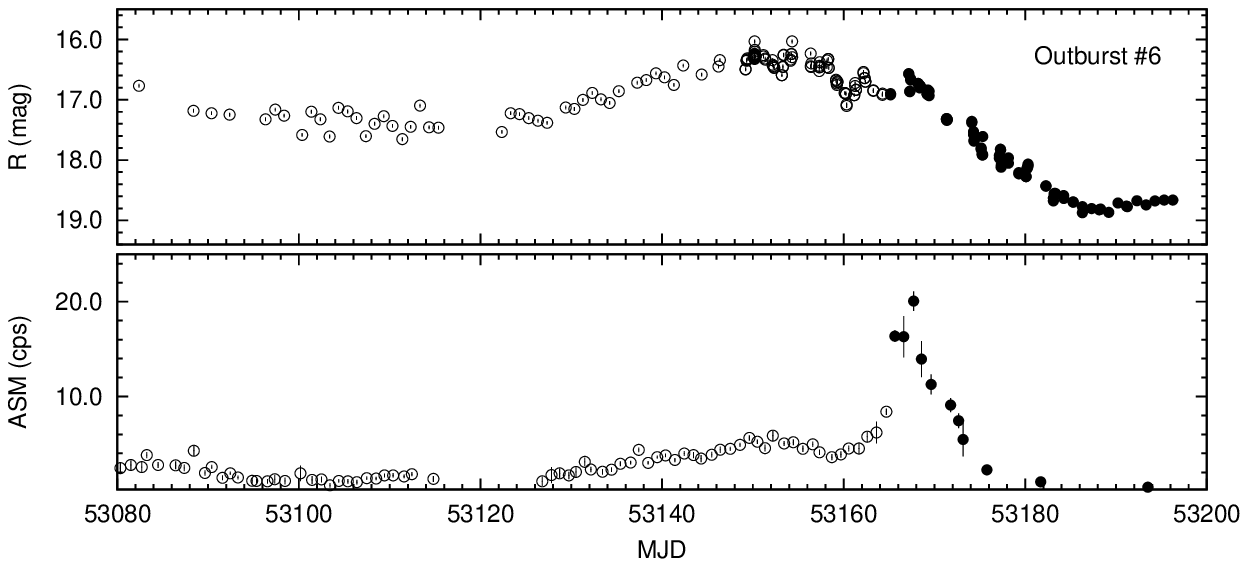}
\caption[Light curve during outbursts of \aql]
{Same as Fig.~\ref{zoom_lc_1-3}, but for outbursts 4--6.
\label{zoom_lc_4-6}}
\end{center}
\end{figure}

\begin{figure}
\begin{center}
\includegraphics{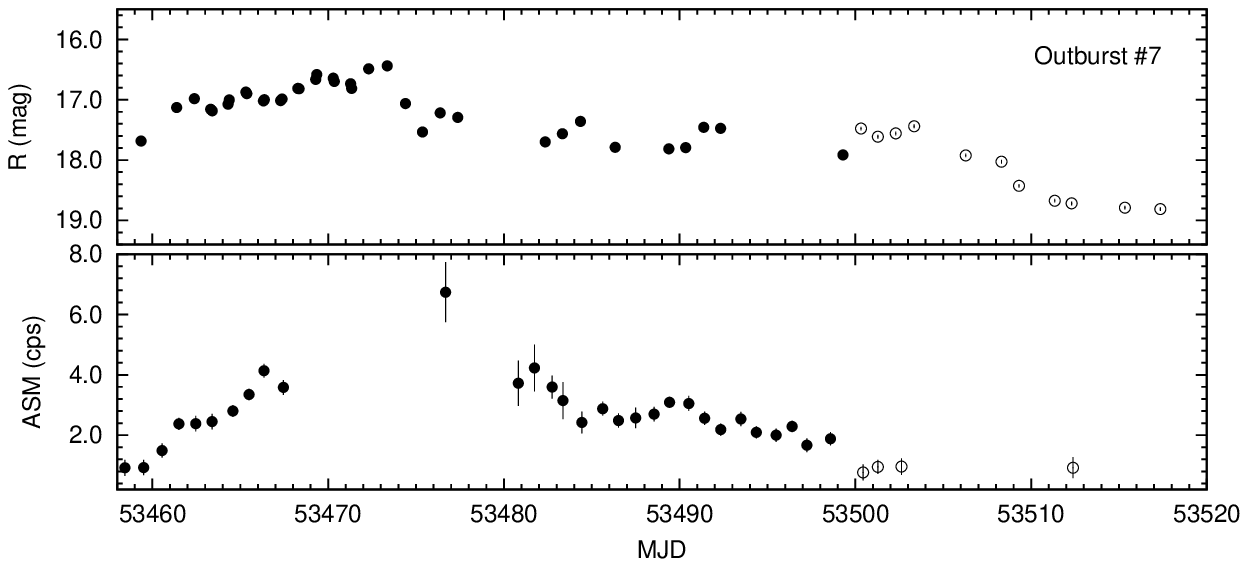}
\includegraphics{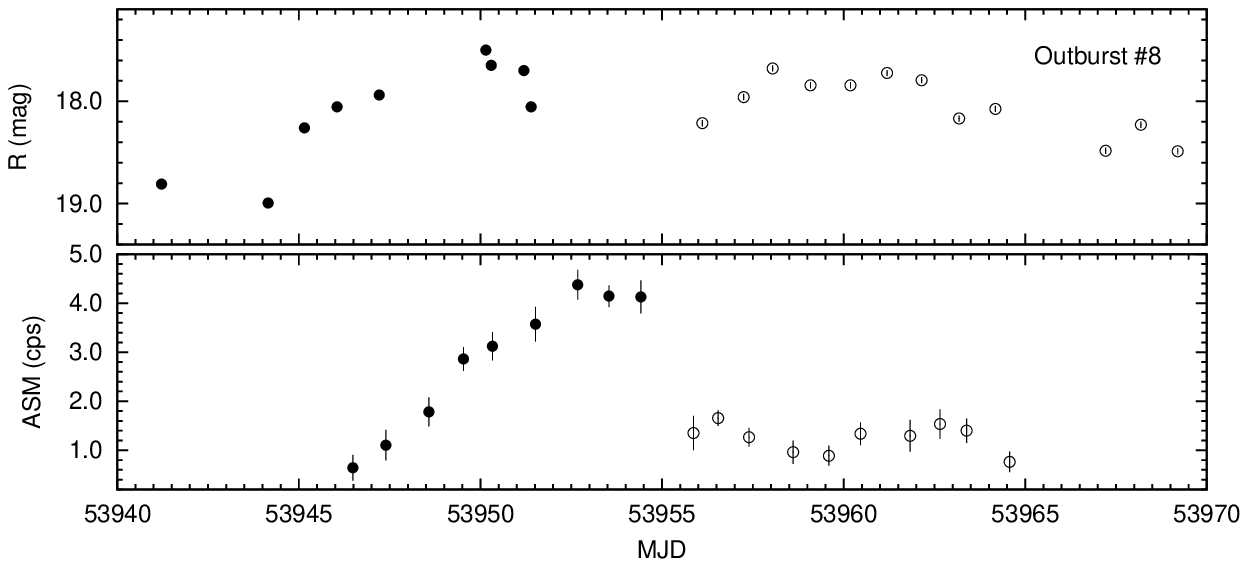}
\includegraphics{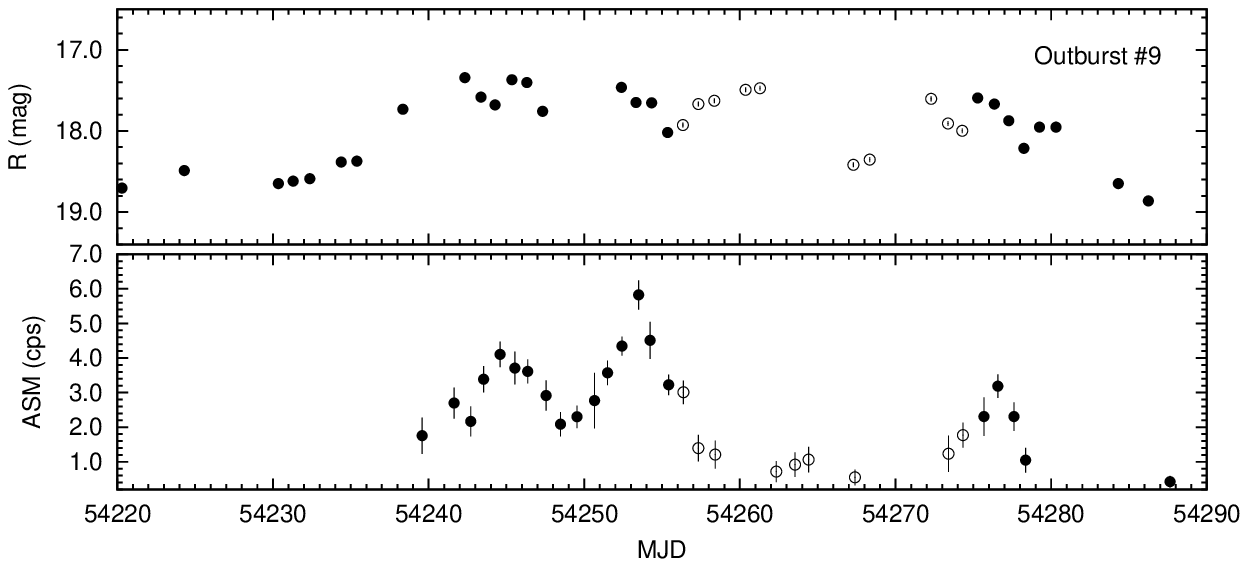}
\caption[Light curve during outbursts of \aql]
{Same as Fig.~\ref{zoom_lc_1-3}, but for outbursts 7--9.
\label{zoom_lc_7-9}}
\end{center}
\end{figure}

\begin{figure}
\begin{center}
\includegraphics{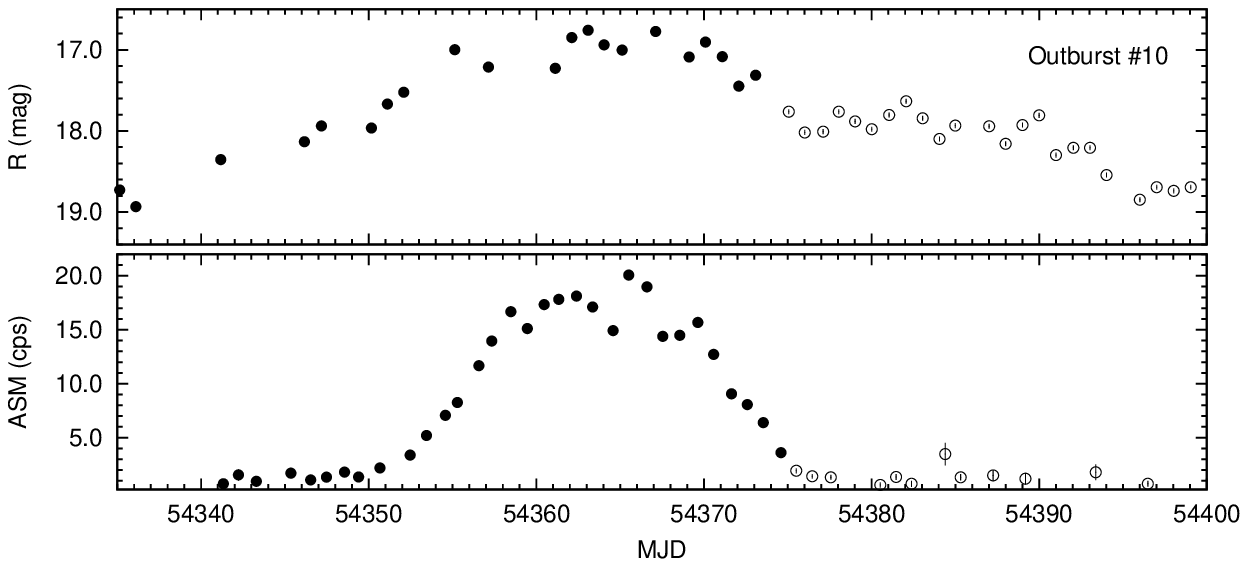}
\caption[Light curve during outbursts of \aql]
{Same as Fig.~\ref{zoom_lc_1-3}, but for outburst 10.
\label{zoom_lc_10}}
\end{center}
\end{figure}

From the long-term OIR monitoring of \aql\ from MJD 50984 -- MJD 54411 
(1998 June 20 -- 2007 Nov 7) only three out of the major OIR outbursts are 
purely FREDs. The remaining are a combination of FREDs with another active 
state where the flux is above quiescence, but does not rise or decay on 
the timescale of a FRED. Instead, a plateau type light curve with 
significant variability on a timescale of days is seen in all bands. 
Sometimes this state persists significantly longer than the FREDs. For 
\aql\ the typical observed FRED rise times are \about\ few days to a week 
and the decay times are \about\ 3--4 weeks. On the other hand, the longer 
lasting, somewhat variable state can last for longer than a month. 
Following \citet{w2002}, who reported observation of such state in the 
neutron star system X1608-52, we call this a ``Low-Intensity State'' 
(LIS). In Fig.~\ref{zoom_lc_1-3} -- Fig.~\ref{zoom_lc_10} we show blowups of
the R-band and ASM light curves for each of the major outbursts in 
Fig.~\ref{longterm} (also tabulated in Table~\ref{tab:outburst_log}). FREDs and
LISs are shown by different symbols in Fig.~\ref{zoom_lc_1-3} -- 
Fig.~\ref{zoom_lc_10}. During a LIS, the OIR flux is near the 
maximum, but always slightly 
lower than the corresponding maximum flux observed during a pure FRED 
outburst. For LISs, the variability in the light curve is more complicated than
the typical `up to a maximum, then down to quiescence' morphology seen for 
FREDs. For a comparable optical brightness, the soft X-ray flux is a 
factor of seven or more lower in a LIS than in a FRED. This high 
optical-to-X-ray flux-ratio, discussed in greater detail in 
\S\ref{sec:r-asm}, is one of the defining characteristics of a LIS. 

\subsection{Optical--soft X-ray correlations\label{sec:r-asm}}

The strongest evidence for the existence of an independent LIS comes from 
optical--X-ray correlations. From the long-term OIR monitoring data, we 
estimated R-band flux-density from \aql\ as described in 
\S\ref{sec:oir-data}.  The RXTE/ASM one day averaged count-rates were 
converted to flux-density as well (\S\ref{sec:asm-data}). Both light 
curves were then searched for quasi-simultaneous optical--X-ray data where 
the observation times in each bandpass were not separated by more than one 
day. In Fig.~\ref{allfluxes} we show the optical--soft-X-ray flux 
correlation during the major outbursts for which we had quasi-simultaneous 
ASM and R-band data.

The optical and soft-X-rays increase approximately linearly with 
$f_{opt}/f_{X-ray}\sim 1.3-1.5$ for FRED outbursts and $\sim 9-12$ for the 
LISs. Given the considerable scatter in the data, we have not attempted to 
fit the data with any function and the flux-ratios quoted above, as shown in 
Fig.~\ref{allfluxes}, are a guide to the eye. For the same X-ray 
brightness, there is no strong hint of hysteresis, i.e. the optical 
brightness during outburst rise is not significantly different from than 
that during the outburst decline. 
The absence of optical--X-ray hysteresis in \aql, goes to 
support the hypothesis that either jets in \ns\ systems are generally 
weaker or the jet break occurs at a lower frequency 
\citep{mf2006,migliari2006,r2007a}.

Both FREDs and LISs start/end the outburst at quiescence and hence have a 
common start/end-point near the lower-left region of $f_{opt}-f_{X-ray}$ 
plane. For some of the weakest outbursts, the ASM fluxes are within a few 
sigma of detection threshold. At such low luminosities, both FREDs and 
LISs occupy similar regions on the $f_{opt}-f_{X-ray}$ plane. It 
has been observed that for some low luminosity outbursts, the source may never 
come out of the X-ray hard state during the entire period of activity
\citep[e.g. see][]{rodriguez2006}. Even 
though the peak optical brightnesses of the brightest LISs are comparable 
to the peak optical brightness during FREDs, LISs are typically fainter 
than FREDs. The distinction between LIS and FRED outbursts stands out 
clearly when optical and X-ray data from {\em all} the major outbursts 
since 1999 June 20 are plotted on $f_{opt}-f_{X-ray}$ plane 
(Fig.~\ref{allfluxes}). LISs and FREDs occupy separate regions of the 
$f_{opt}-f_{X-ray}$ plane and evolve on different timescales, suggesting 
different physical mechanisms for their origin.

\section{OIR color evolution\label{sec:colors}} 

In Fig.~\ref{r-j_1} and Fig.~\ref{r-j_2} the time evolution of the R and 
J-band magnitudes and 
that of the R$-$J color is shown for the major outbursts that were observed 
in both bandpasses. There seems to be no significant inter-band time lag 
during rise or decay of an outburst.  
Irrespective of the light curve morphology during the outburst, the R$-$J 
color is always bluer during an outburst compared to the quiescent value, 
and therefore $(R-J)_{outburst} < (R-J)_{quiescence}$ as seen in 
Fig.~\ref{r-j_1} and Fig.~\ref{r-j_2}.
This decrease in R$-$J color 
(becoming bluer) is characteristic of increased irradiation heating of the 
outer disk. 

During quiescence, the disk flux is very low compared to the 
combined flux from the contaminating star and the donor star. It is however 
interesting to note that the R$-$J color during an LIS is significantly 
redder than that during the peak of a FRED outburst.

Fig.~\ref{aql-cmd1} and Fig.~\ref{aql-cmd2} shows the OIR \cmd\ for the 
different outbursts. The 
R-band magnitude is plotted along the ordinate. As is typical for 
conventional \cmd s, the ordinate axis has been reversed so that higher 
luminosities (i.e. lower magnitude) are plotted near the top and 
vice-versa.  The R$-$J color is plotted along the abscissa.

During periods of quiescence, the source flux is low and essentially 
dominated by photons from the contaminating star and the donor, and it lingers 
near the bottom-right corner of the \cmd. As the source enters an 
outburst, the brightness increases and the color becomes bluer, due to an 
increasing contribution of the accretion disk flux. 
In Fig.~\ref{aql-cmd1} and Fig.~\ref{aql-cmd2}, the observations when the 
source was in the hard 
state during the rise of an outburst, when it was in the hard state during 
the decay and when it was in the soft state are shown by different 
symbols. 
As noted in \S~\ref{sec:r-asm}, there are also outbursts
when the source spent the entire period of activity in an X-ray hard 
state \citep{rodriguez2006}. The OIR data during such outbursts are shown by
open squares. 
There seems to be no significant change in OIR color across the 
X-ray spectral state transitions.

We attempted to test the simple ansatz that the observed evolution on the 
\cmd\ is consistent with the thermal heating of a \bb. Our goal is to see 
whether the flux and color variations are correlated as would be expected 
if the only change is in the disk temperature.  For simplicity, and 
because we had only two band-passes, we considered the entire annulus to 
be a single temperature blackbody, even though real accretion disks are 
likely to have a significant temperature gradient. In this simple model, 
the R$-$J color is determined by the temperature of the annulus and the 
R-band brightness is proportional to the projected area of the annulus and 
inversely proportional to the square of the distance to the source. Since 
the observed brightness depends on the projected area, we multiplied the 
area of a disk of radius $R_{disk}$, by a multiplicative factor $f<1$, and 
used this area as the area of the annulus. Thus in this simple model the 
brightness is proportional to $f(R_{disk}/D_{disk})^2cos(i)$, where 
$D_{disk}$ and $i$ are the source distance and $i$ is the orbital 
inclination respectively. The only adjustable parameter here is $f$, which 
changes the overall projected area of the annulus. The parameter $f$ is 
changed from outburst to outburst, keeping other parameters constant. 
Fitting the ellipsoidal variations observed in the quiescent light curve, 
\citet{rwy2001} estimated that $36\degr\leq i\leq55\degr$. We assumed an 
inclination of $45\degr$. An upper limit on $R_{disk}$ can be presumed to 
be equal to the radius of the Roche lobe of the accretor and therefore can 
be estimated from the mass ratio and orbital period \citep{e1983}. 
Evolution of the source on the \cmd\ was generated by calculating the 
color and brightness of the annulus at varying temperature. We note that 
the track generated by the heating/cooling of this constant area blackbody 
broadly describes the evolution of the source during its outbursts.
For a purely viscous disk (no irradiation) the spectral index between R and
J bands (as defined in Eq.~\ref{eq:color-slope}) should lie between $1/3$ 
and $2$. The observed range in $R-J$ color during outbursts is approximately
between $1.35$ and $2.1$, which using Eq.~\ref{eq:color-slope} translates 
to a spectral slope between $0.27$ and $-0.95$ respectively. The observed 
range in spectral indices is quite different from what could be expected from 
a purely viscous disk and favors irradiation heating.

We found that even for the largest possible disk radius of \about 5 light 
seconds ($M_{donor}=0.5\msun$, $M_{accretor}=1.5\msun$, $P_{orb}=18.95$ 
hour), the model-predicted R-band flux is too small for the lowest 
distance estimate of 4 kpc by \citet{r2001}. Using $D_{disk}=3$ kpc seems 
to lie close to the observed datapoints (see Fig.~\ref{aql-cmd1} and 
Fig.~\ref{aql-cmd2}). Since we 
are not doing detailed spectral fits, we do not claim the distance to the 
source to be 3 kpc, but show that this very simple model can roughly 
predict the outburst behavior of the accretion disk. The model curves 
overplotted in Fig.~\ref{aql-cmd1} and Fig.~\ref{aql-cmd2} are not fits to the 
datapoints but 
created using a disk of radius 5 light-seconds, inclined at an angle of 
45$\degr$ to the observer and at a distance of 3 kpc from the observer, 
viz. a set of parameters that is generally consistent with the observed 
values, and with a small range (0.6-0.8) in the value of the 
multiplicative factor $f$. The general consistency of the data over many 
outbursts supports the idea that the OIR flux is largely due to thermal 
emission from an irradiated disk. This is in contrast to recent 
observations of accreting black hole systems in which a strongly 
non-thermal component is often observed in the near IR 
\citep[][ Russell, Maitra \& Fender, in prep.]{bb2004,h2005,r2006}.

\section{Discussion\label{sec:discuss}}

\begin{figure} 
\begin{center}
\includegraphics[width=0.90\textwidth, angle=0]{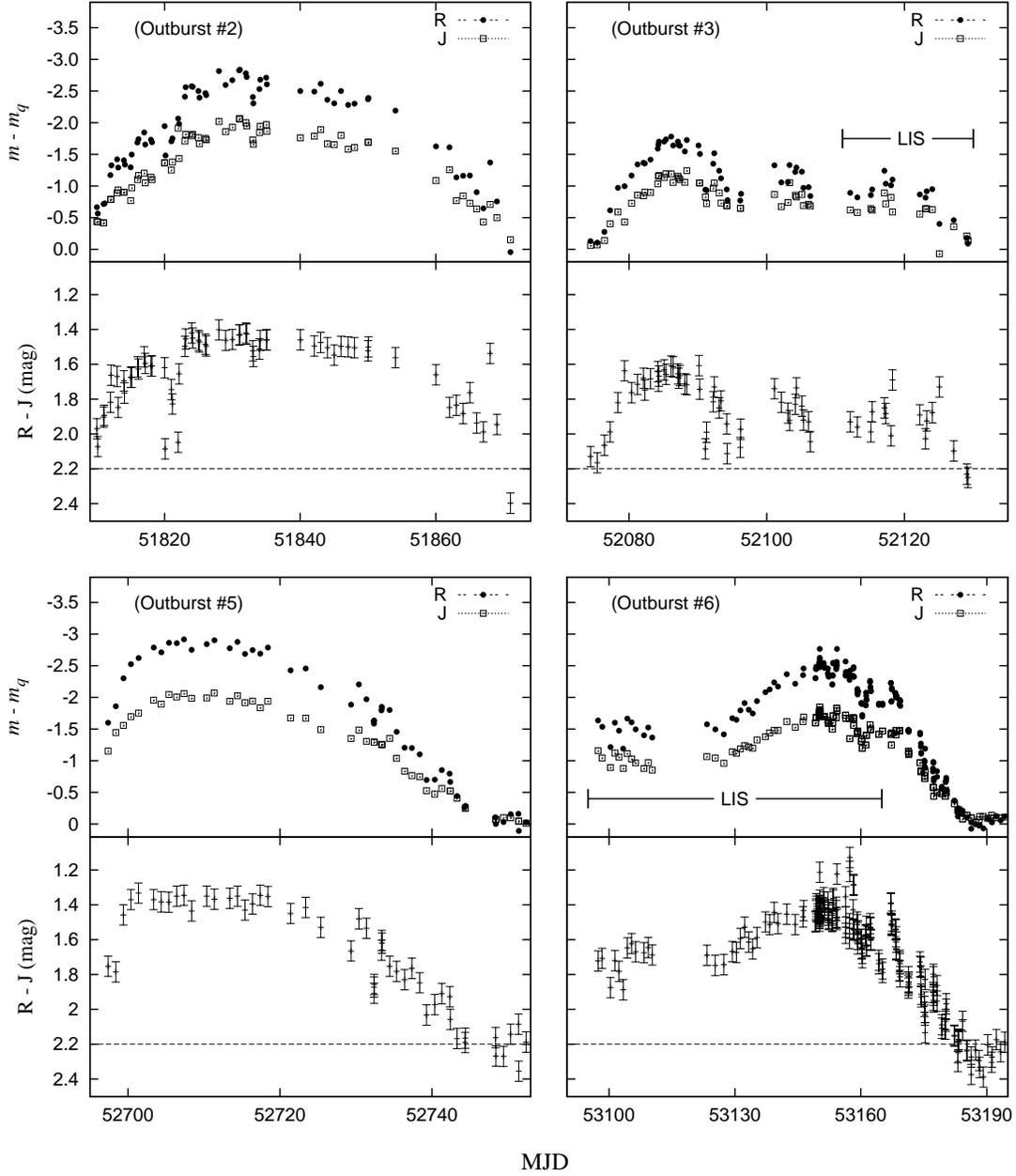}
\caption[Evolution of J-R color during various outbursts of \aql]
{Evolution of J-R color during outbursts 2, 3, 5 and 6 of \aql. The
increase in brightness from quiescence is shown in the top panels, with the 
R-band data shown by filled circles and J-band data shown by open
squares. The outburst serial number (Table.~\ref{tab:outburst_log})  is
given in parentheses. Variation of corresponding J-R color is shown in the
bottom panels. Epochs of LIS as inferred from 
Figs.~\ref{longterm}--\ref{zoom_lc_10} are labelled. Unless labelled as a LIS,
the OIR/soft-X-ray light curve morphology resembles that of FRED outbursts.
\label{r-j_1}} 
\end{center}
\end{figure}

\begin{figure} 
\begin{center}
\includegraphics[width=0.90\textwidth, angle=0]{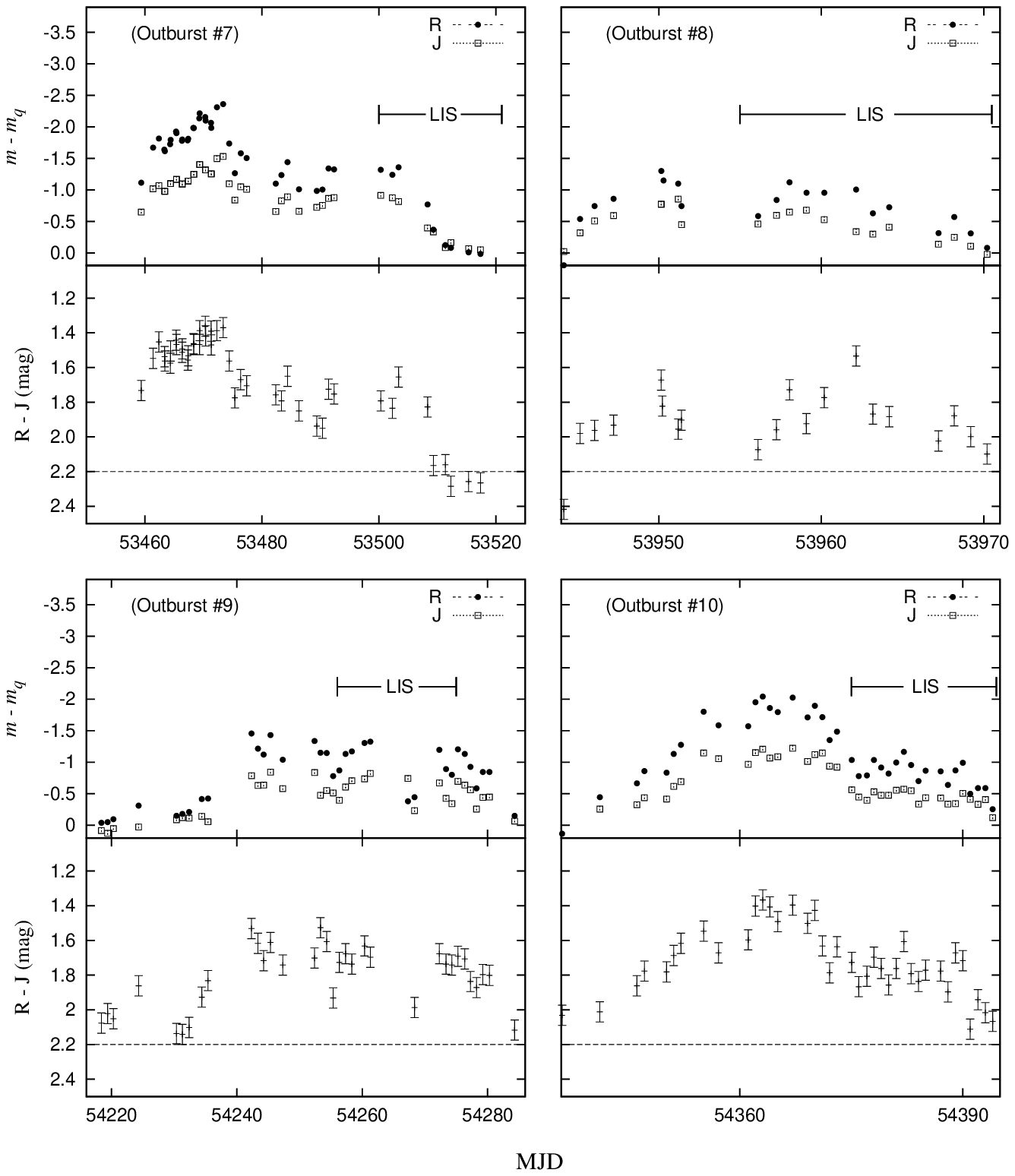}
\caption[Evolution of J-R color during various outbursts of \aql]
{Same as Fig.~\ref{r-j_1}, but for outbursts 7--10.
\label{r-j_2}} 
\end{center}
\end{figure}

Multi-wavelength monitoring of the neutron star soft X-ray transient \aql\ 
in near-IR, optical and soft X-ray data from RXTE/ASM over the past nine 
years has allowed us to study the evolution of the source with 
unprecedented coverage. The OIR and X-ray light curves show a wide range 
of outburst light curve patterns. However the major outbursts show two 
distinct light curve morphologies. The {\em fast-rise and 
exponential-decay} \citep[FRED;][]{csl1997} type outbursts characteristic 
of many soft X-ray transients is seen in \aql\ too. However, only three of 
the major outbursts seen since 1998 June show just a FRED type light curve 
profile. More often a prolonged, somewhat variable ``Low-Intensity state'' 
\citep[LIS;][]{w2002}, uniquely characterized by an extremely high 
optical-to-soft X-ray flux-ratio, is seen either before or after a FRED.
It has been observed that for some of these ``soft X-ray faint'' LIS 
outbursts, \aql\ never comes out of the X-ray hard state during the entire 
outburst \citep{rodriguez2006}.
Other than these major outbursts, small-scale flaring activities were also 
observed. While the origin of such small-scale short activity remains 
unknown, such activity has been observed in other sources like XTE 
J1550-564 by \citet{ss2005} who suggested that these small outbursts could 
be due to discrete accretion events. We suggest the following framework 
for understanding these results:

  {\em Irradiation dominates OIR flux for both FRED and LIS
	  outbursts of \aql.}
  During the course of FRED and LIS outbursts of \aql\, the evolution of
  the source on the OIR \cmd\ is consistent with heating/cooling of an 
  irradiated outer accretion disk as a constant area \bb.
  A classical viscous accretion disk (e.g. a Shakura-Sunyaev disk) spectrum 
  has a frequency dependence of $F_{\nu}\sim \nu^{1/3}$ between 
  $kT(R_{out})/h \ll \nu \ll kT(R_{in})/h$ where  $T(R_{in})$ and $T(R_{out})$
  are the disc temperatures at the inner and outer edges of the disk. At 
  frequencies much smaller than $kT(R_{out})/h$ the disk spectrum is simply the
  Rayleigh-Jeans region of a \bb\ with $F_{\nu}\sim \nu^2$. For soft X-ray
  transients $kT(R_{in})/h$ is
  typically in far-UV or soft X-rays. Therefore in our 
  case the flux ratios (a.k.a. colors) of two bands, from a purely viscous 
  disc are expected to be constant as long as {\em both bands} are either on 
  the viscous ($\nu^{1/3}$) or Rayleigh-Jeans ($\nu^2$) regime. In the special 
  case when the break frequency between the viscous dissipation and the 
  Rayleigh-Jeans regime given by $\nu_{break}=kT(R_{out})/h$ lies between R
  and J bands, the spectral index should lie between $1/3$ and $2$.
  The observed systematic and continuous change in color (and the corresponding 
  range in the spectral index from $-0.95$ to $0.27$; \S\ref{sec:colors}) seen 
  in Fig.~\ref{r-j_1} and Fig.~\ref{r-j_2} is strongly 
  suggestive of an irradiative origin for the observed OIR emission and not 
  of a purely viscous origin. 

\begin{figure}
\begin{center}
\includegraphics[width=0.73\textwidth, angle=-90]{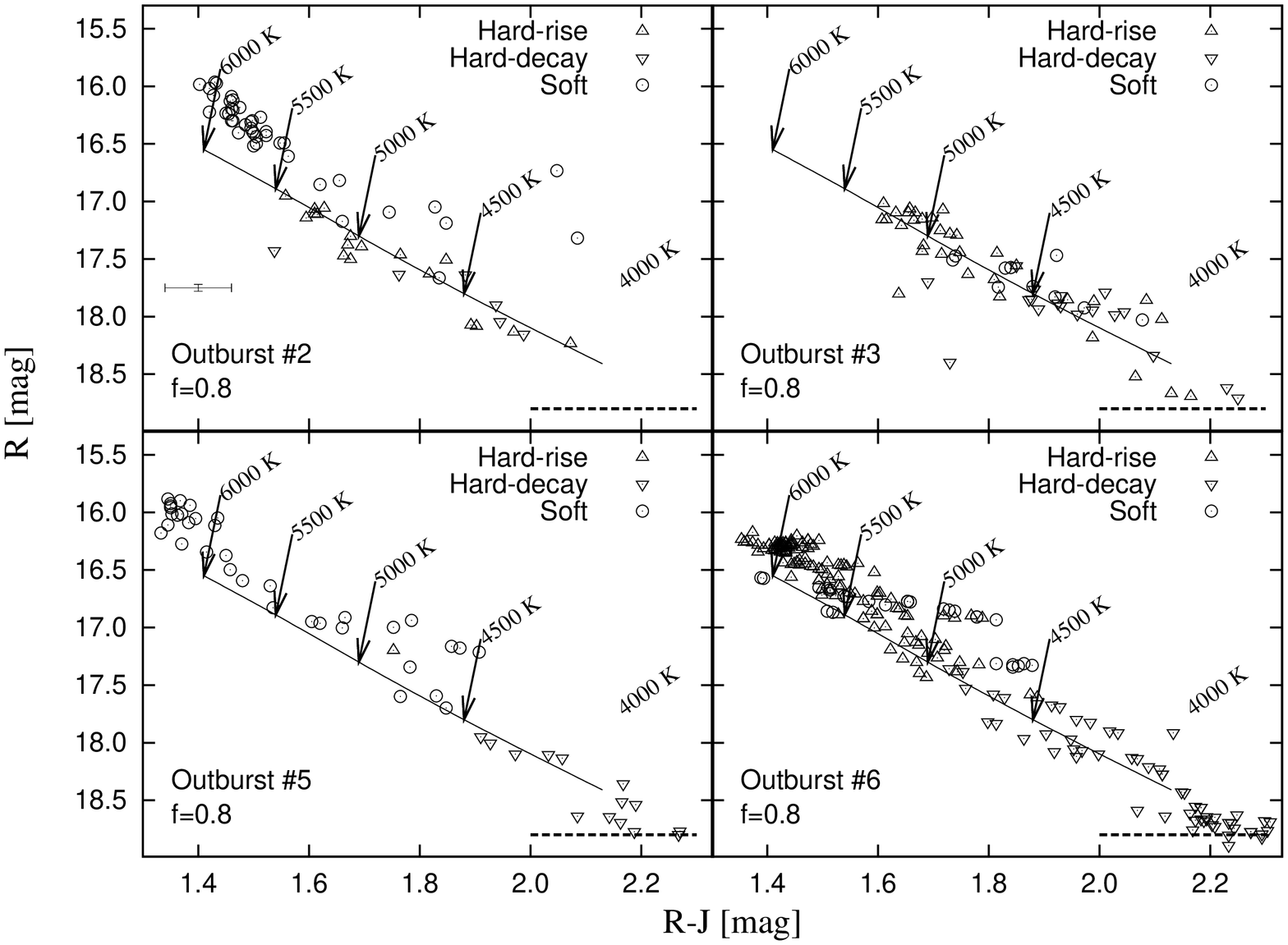}
\caption[Evolution of the outbursts on OIR \cmd]
{Evolution of outbursts 2, 3, 5 and 6 of \aql\ on the OIR \cmd. 
Observations when the source was in hard/power law state during the rise 
of an outburst are shown by ($\bigtriangleup$). The decaying hard state, 
are shown by ($\bigtriangledown$). Outbursts where state transitions were 
not reported are shown by ($\sq$). Observations during soft/thermal 
dominated state are shown by ($\circ$). The quiescent level is shown by the 
dashed line. 
A typical error bar is shown in the top left panel near (1.4, 17.75). 
The solid, 
curve from bottom-right to top-left in each panel shows the 
color and brightness expected by heating a disk-shaped \bb\ of radius 5 
light seconds, inclined to the observer at $45\degr$ and at a distance of 3
kpc. The quantity $f$ represents a multiplicative area correction factor.
The expected color and magnitude of the disk between temperatures of 
4000 --6000 K are labelled. 
\label{aql-cmd1}} 
\end{center}
\end{figure}

\begin{figure}
\begin{center}
\includegraphics[width=0.73\textwidth, angle=-90]{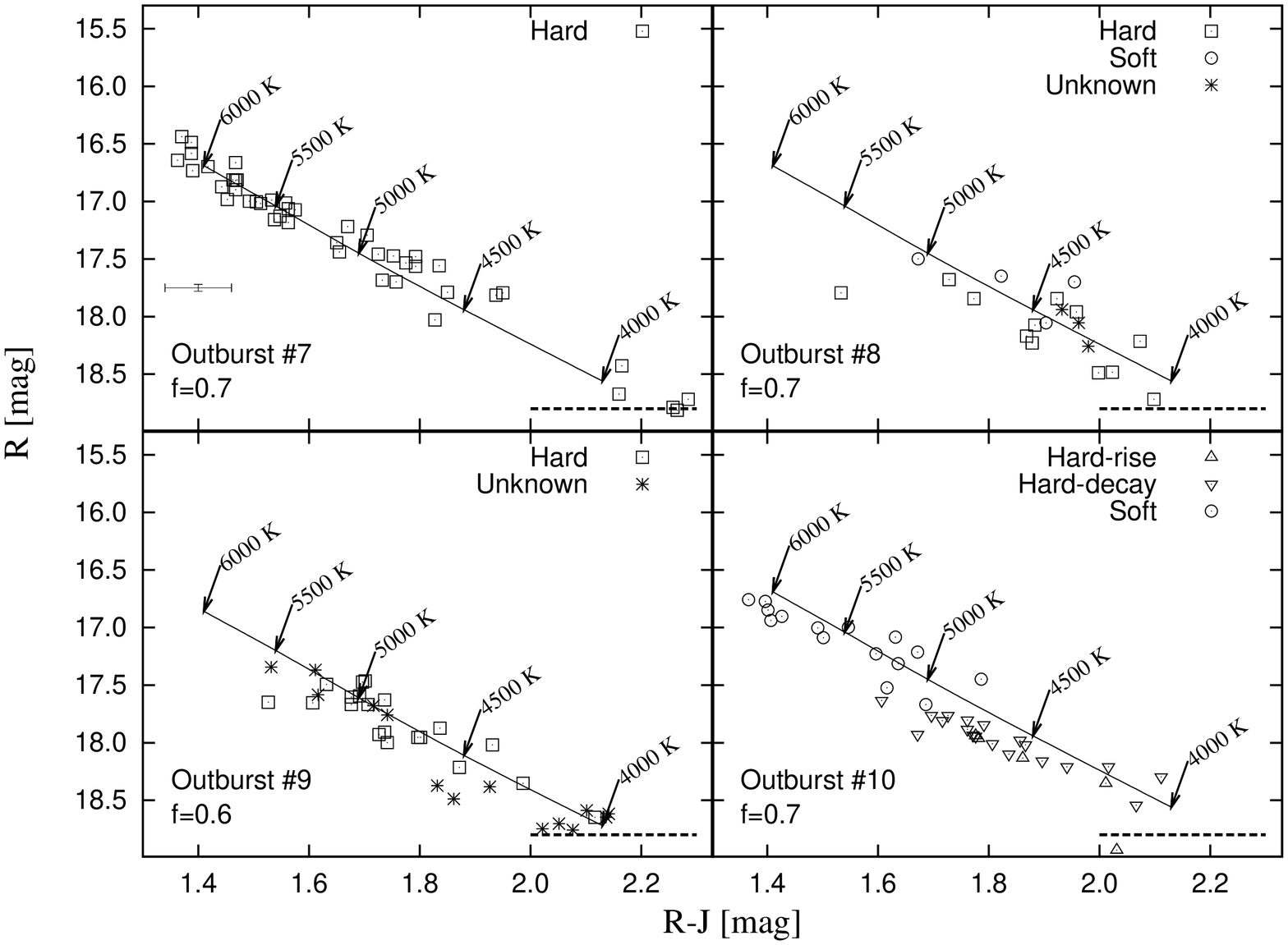}
\caption[Evolution of the outbursts on OIR \cmd]
{Same as Fig.~\ref{aql-cmd1}, but for outbursts 7, 8, 9 and 10.
\label{aql-cmd2}} 
\end{center}
\end{figure}

  Neither the OIR color nor its brightness change sharply during an
  X-ray spectral state transition. This confirms that for \aql\, the outer
  accretion disk, which emits most of the thermal OIR radiation, is not
  affected by X-ray spectral state transitions which change the physical
  environment of the inner regions of the accretion disk. This is in 
  sharp contrast with \bh\ binaries like 4U 1543-47, GX 339-4 
  \citep{bb2004,h2005} where near IR colors are strongly correlated with
  X-ray state changes and the correlation interpreted as evidence for 
  non-thermal emission.
  In FREDs, a slow increase in brightness above that 
  predicted by our constant area \bb\ model, as the source becomes bluer 
  and brighter, is most likely due to increase in the surface area
  of the disk (e.g most prominent for outburst \#2 and \#5 in 
  Fig.~\ref{aql-cmd1}). It is possible that at highest mass accretion rates 
  the high X-ray flux from the central 
  inner disk causes the flaring of the outer disk and hence increase the 
  effective surface area. In contrast, during the LISs, when the source 
  probably spends the whole period of activity in hard X-ray state, there 
  does not seem to be any significant change in the disk area (e.g. outburst
  \#7 in Fig.~\ref{aql-cmd2} ). 
  Therefore it is likely that the effect of X-ray state transition on the outer 
  accretion disk is both an increase of the outer disk temperature due to 
  enhanced irradiation from the inner disk and 
  an increase in the emitting area of the outer disk due to flaring. 
  Total irradiation (hence OIR flux) is a bit smaller in
  LIS than in peak FRED. This could be due either to lower \mdot, or 
  to radiative inefficiency lowering $L/\mdot$ ratio.

\begin{figure}
\begin{center}
\plotone{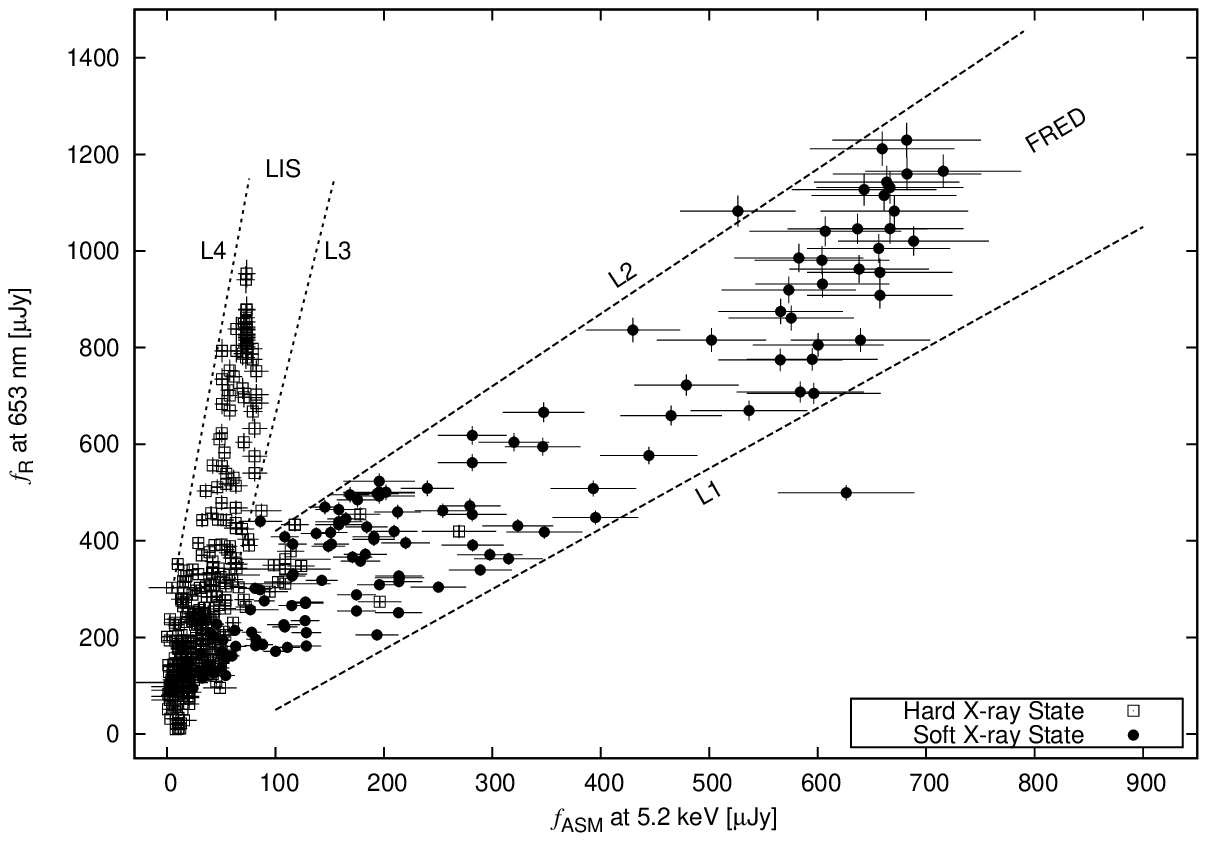}
\caption[Optical--soft-X-ray correlation: All outbursts combined]
{Optical--soft-X-ray correlation in \aql. Data from all the major outbursts
from MJD 50984 -- MJD 54411 (1998 June 20 -- 2007 Nov 7) 
for which the X-ray state could be determined from contemporaneous 
RXTE observations \citep[][also Manuel Linares, priv. comm.]
{mc2003a,mc2003b,rks2004,mb2004,rodriguez2006,y2007}. 
The dashed lines L1 and L2 which follow the equations $f_R=1.25f_{ASM} - 75$ 
and $f_R=1.5f_{ASM} + 270$ respectively, roughly form an envelope containing 
most of the FRED points.
The dotted lines L3 and L4 which follow the equations 
$f_R = 9f_{ASM} - 240$ and $f_R = 12f_{ASM} + 240$ respectively, roughly form
an envelope containing most of the LIS points.
Filled circles represent data taken during {\em soft} X-ray state and open
squares are those during {\em hard} X-ray state.
\label{allfluxes}}
\end{center}
\end{figure}

  {\em Broad band quasi-simultaneous observations in LIS suggest 
	  outflow.} 
  Soft X-ray flux is very low in LIS, but hard X-rays are high 
  \citep[][Linares et al. in preparation,]{atel259}.
  While absence of soft, thermal X-ray photons suggests that the inner edge 
  of the accretion disk is cold and probably receded during the LIS,
  the high OIR brightness (see Fig.~\ref{allfluxes} for the dramatic 
  difference in $f_{opt}/f_{X-ray}$ between LIS and FRED outbursts )
  as well as type I X-ray bursts suggests that
  a significant amount of mass is being accreted by the outer accretion disk
  and fed to the inner region.

  Radio detection of the source during LIS \citep{atel286} with 
  flat spectral slope hints toward a jet outflow. 
  However, since the evolution of the source on OIR \cmd\ is consistent 
  with irradiation, it can be concluded that the jet, if present, is either 
  quite weak in 
  terms of the input power, or the transition from optically thick 
  to optically thin jet emission occurs at a wavelength much longer than 
  J-band. 
  Current models for jets from compact systems suggest that this might imply
  a large jet launching area or the jet 
  base, or much weaker particle acceleration \citep{mff2001}. Thus our 
  observations support the current hypothesis that jet-luminosity is not 
  significant in most neutron star X-ray binary systems \citep{mf2006,f2006}.

  {\em A possible scenario for the LIS might be that an \mdot\ 
  that would ordinarily be high enough to move the disk inwards, and thus 
  trigger a soft state, somehow doesn't do that, and launches an outflow 
  instead.}
  Thus the standard
  processes associated with the DIM are disrupted by the absence
  of an inner disk in this high \mdot\ state. If the accretion flow in 
  the inner region in a LIS is radiatively inefficient, it can drive 
  the observed outflow, as suggested by recent theoretical works 
  \citep[see e.g.][]{bb1999,m2001,n2005}.

\acknowledgments

It is a pleasure to thank the SMARTS observers J. Espinoza and D. Gonzalez 
for taking the data over the years, and R. Winnick, J.Nelan and M. Buxton 
who accommodated our many requests to revise the observing schedule. 
We would like to thank Ron Remillard of the ASM team for his help in 
converting ASM count-rates to flux-density.
DM would 
like to thank Manu Linares for providing PCA colors of all the RXTE 
observations of \aql\ ahead of publication, and Sera Markoff, Dave Russell and 
Rob Fender for discussion and useful comments. We would like to thank the 
anonymous referee for constructive criticisms which improved the paper.
DM is supported by Netherlands Organisation for Scientific 
Research (NWO) grant number 614000530. CDB acknowledges support 
from the National Science Fopundation grant NSF-AST 0707627. This work has 
made use of the Astrophysics Data System Abstract, the MIT/ASM data 
extraction website. This work has also used data products from the Two 
Micron All Sky Survey, which is a joint project of the University of 
Massachusetts and the Infrared Processing and Analysis Center/California 
Institute of Technology, funded by the National Aeronautics and Space 
Administration and the National Science Foundation.

{\it Facilities:} \facility{RXTE}, \facility{SMARTS}

\clearpage

\begin{deluxetable}{ccclc}
\tablecolumns{5}
\tablewidth{0pc}
\tablecaption{Date of peak flux during the outbursts of \aql\ between
	MJD 50984 -- MJD 54411 (1998 June 20 -- 2007 Nov 7).
	\label{tab:outburst_log}}
\tablehead{
	\colhead{R-band peak} & \colhead{J-band peak} & \colhead{ASM peak} &
	\colhead{Outburst type: Evolution of} & \colhead{Major Outburst} \\
	\colhead{(MJD)} & \colhead{(MJD)} & \colhead{(MJD)} &
	\colhead{light curve morphology} & \colhead{Number}
}

\startdata
 51076 & --    & --    & Mini				 & -- \\
 51320 & --    & 51321 & Major: FRED $\rightarrow$ LIS	 & 1  \\
 51676 & --    & --    & Mini				 & -- \\
 51730 & --    & --    & Mini				 & -- \\
 51831 & 51831 & 51837 & Major: FRED			 & 2  \\
 52086 & 52088 & 52087 & Major: FRED $\rightarrow$ LIS	 & 3  \\
 52192 & 52192 & --    & Mini				 & -- \\
 52226 & 52226 & --    & Mini				 & -- \\
 52338 & --    & 52336 & Major: FRED			 & 4  \\
 52707 & 52711 & 52705 & Major: FRED			 & 5  \\
 52923 & 52923 & --    & Mini				 & -- \\
 53150 & 53150 & 53167 & Major: LIS  $\rightarrow$ FRED  & 6  \\
 53268 & 53268 & --    & Mini				 & -- \\
 53473 & 53473 & 53476 & Major: FRED $\rightarrow$ LIS	 & 7  \\
 53579 & 53580 & --    & Mini				 & -- \\
 53950 & 53951 & 53952 & Major: FRED $\rightarrow$ LIS	 & 8  \\
 54242 & 54245 & 54253 & Major: FRED $\rightarrow$ LIS	 & 9  \\
 54363 & 54367 & 54365 & Major: FRED $\rightarrow$ LIS	 & 10
\enddata
\end{deluxetable}


\clearpage



\end{document}